\documentclass[conference]{IEEEtran}
\IEEEoverridecommandlockouts
\usepackage{cite}
\usepackage{amsmath,amssymb,amsfonts}
\usepackage{graphicx}
\usepackage{textcomp}
\usepackage{caption}
\usepackage{subcaption}
\usepackage{xcolor}
\usepackage[ruled,vlined]{algorithm2e}
\def\BibTeX{{\rm B\kern-.05em{\sc i\kern-.025em b}\kern-.08em
    T\kern-.1667em\lower.7ex\hbox{E}\kern-.125emX}}
\begin{document}

\newtheorem{proposition}{Proposition}

\title{Hybrid Vehicular and Cloud Distributed Computing: A Case for Cooperative Perception}
%\thanks{This work was supported in part by the CONIX Research Center, one of six centers in JUMP, a Semiconductor Research Corporation (SRC) program sponsored by DARPA.}

\author{
    \IEEEauthorblockN{Enes~Krijestorac\IEEEauthorrefmark{1}, Agon Memedi\IEEEauthorrefmark{1}\IEEEauthorrefmark{3}, Takamasa Higuchi\IEEEauthorrefmark{2}, Seyhan Ucar\IEEEauthorrefmark{2}, Onur Altintas\IEEEauthorrefmark{2}, Danijela Cabric\IEEEauthorrefmark{1}}
    \IEEEauthorblockA{\IEEEauthorrefmark{1}\textit{Electrical and Computer Engineering Department,} \textit{University of California, Los Angeles},
    USA\\
    \IEEEauthorrefmark{3}\textit{Heinz Nixdorf Institute and Dept. of Computer Science, Paderborn University}, Germany
    \\\ enesk@ucla.com, memedi@ccs-labs.org, danijela@ee.ucla.edu} 
    \IEEEauthorblockA{\IEEEauthorrefmark{2}\textit{InfoTech Labs, Toyota Motor North America R\&D}, \textit{Mountain View, California}, USA
    \\ \{takamasa.higuchi, seyhan.ucar, onur.altintas\}@toyota.com}
}

\maketitle
\newcommand{\pp}[1]{\textcolor{blue}{#1}}   
\begin{abstract}
%Cooperative vehicular perception systems seek to expand a vehicle's field of view by enabling cars to share the data from their environment perception sensors via wireless communication and thereby allowing vehicles to achieve a global view of the traffic environment. 
%Computationally demanding and time-sensitive vehicular applications are often bottlenecked by the computing power of individual vehicles. 
%One of the main bottlenecks of cooperative perception is the limited computing power of individual cars. 
In this work, we propose the use of hybrid offloading of computing tasks simultaneously to edge servers (vertical offloading) via LTE communication and to nearby cars (horizontal offloading) via V2V communication, in order to increase the rate at which tasks are processed compared to local processing. 
Our main contribution is an optimized resource assignment and scheduling framework for hybrid offloading of computing tasks. The framework optimally utilizes the computational resources in the edge and in the micro cloud, while taking into account communication constraints and task requirements. While cooperative perception is the primary use case of our framework, the framework is applicable to other cooperative vehicular applications with high computing demand and significant transmission overhead. The framework is tested in a simulated environment built on top of car traces and communication rates exported from the Veins vehicular networking simulator. We observe a significant increase in the processing rate of cooperative perception sensor frames when hybrid offloading with optimized resource assignment is adopted. Furthermore, the processing rate increases with V2V connectivity as more computing tasks can be offloaded horizontally. 
%While cooperative perception is used to analyze the performance of hybrid offloading, the framework is applicable to other applications with cooperative tasks and high computing demands.

\end{abstract}


\section{Introduction}
%\IEEEPARstart{A}{utonomous} cars and driving assistance systems rely on sophisticated sensors such as radars, lidars and cameras to understand the surrounding environment and act upon sensory feedback. 
%\IEEEPARstart{T}{ogether} with the advances in artificial intelligence, new developments in sensing technology are necessary to achieve autonomous driving systems that surpass human capabilities. 
%However, no matter how advanced perception sensors should become, the perceivable area of a single vehicle is limited. 
\IEEEPARstart{C}{ooperative} vehicular perception systems seek to expand a vehicle's field of view by enabling cars to share the data from their environment perception sensors via wireless communication and thereby allowing vehicles to achieve a global view of the traffic environment. %Different types of messages are exchanged in these systems, including but no limited to: laser scans, raw and compressed images, point clouds and lists of objects abstracted from the raw sensor data \cite{kim2014multivehicle, qiu2018avr}. 
%One of the main bottlenecks of cooperative perception is the computing power of the cars' microprocessors. 
%While a reliable communication channel is essential for the transmission of the perception data, cooperative perception also relies on processing of the sensor data at a very high rate. 
However, granted a high speed wireless communication channel, experimental implementation studies of cooperative perception systems report that the main bottleneck in the frame rate at which cooperative perception systems can operate is the vehicular computational power \cite{qiu2018avr}. 
%The limited rate at which frames can be processed and exchanged presents a major drawback for cooperative perception systems in time sensitive scenarios. If the interval between successive arrival of frames is significant, the objects in the surrounding environment might have changed the position, direction and speed significantly during the inter-arrival period. 
%It is imperative to increase the frame processing rate in cooperative perception systems in order to make them a more compelling technology for use in driving assistance and autonomous driving systems. 

Cooperative perception and similar computationally demanding and time-sensitive vehicular applications can benefit from additional computing power. 
%The limited amount of computing resources in individual cars can be supplemented by leveraging the computing resources in the edge and in neighboring vehicles. 
Edge computing is a promising paradigm where computing resources are placed in close proximity of end users, which are usually mobile. Vehicles can offload some of the computational tasks to edge servers which they can reach {via an LTE connection}. The results of the processed tasks are then be sent back from the edge servers to the interested cars on the road. 
We refer to this type of offloading as \emph{vertical offloading}. 
The vehicle cloudification framework, a paradigm that involves forming virtual cloud servers from vehicles in proximity of each other on the road, can enable \emph{horizontal offloading}. In horizontal offloading, the computing resources and the coordination of the vehicular micro cloud can be utilized for task offloading via V2V communication.
The feasibility of forming stable vehicular micro clouds on the road has been confirmed in \cite{higuchi2017feasibility}.

%Both vertical and horizontal task offloading incurs a transmission delay due to transmission of sensor frames from the source to where they are processed and due to transmission of the results to the receivers. 

In addition to the rate at which sensor frames are processed, there are other considerations that need to be made when it comes to cooperative perception systems. {Firstly, the processing delay of a frame must not be too long, otherwise the processing results would be out of date relative to the actual state of the traffic environment. When offloading tasks, the processing delay includes data frame transmission delay and computing delay.} {Secondly}, we need to assume that there is a limit on the amount of data that can be transmitted over the cellular and V2V links. For the cellular data exchange, the practical reason for this is that there is normally a monetary cost related to the utilization of the cellular link and that there might also be a limit on how much the cellular providers will allow the network to be loaded by the transmission of vehicular perception data, since it could negatively impact the quality of service for other cellular users. Likewise, there is a practical limit on how much data transfer can occur over V2V links, since the data transfer of cooperative perception data could cause congestion and disrupt other services that rely on the shared V2V channel.

Approaches to offloading of vehicular computing tasks to the edge or to the nearby cars have been proposed in the literature before. A framework for task allocation in horizontal offloading was proposed in \cite{hattab2019optimized}. However, in this work, the delay due to transmission of the computing task data has not been considered, therefore this particular framework cannot be applied on cooperative perception or other computing tasks that would have a large transmission overhead when offloaded. Another comprehensive framework for task offloading was developed in \cite{feng2017ave}. While \cite{feng2017ave} considers the data transfer of computing tasks, it focuses only on horizontal task offloading. A mechanism for both horizontal and vertical task offloading was proposed in \cite{zhu2018fog}. While the system proposed in \cite{zhu2018fog} could in theory be applied to cooperative perception or other data intensive computing tasks, it does not give any consideration to cellular/V2V traffic overhead limitations.   

{In this paper, we propose \emph{hybrid horizontal and vertical offloading} of time-sensitive cooperative perception computing tasks with the goal of increasing the rate at which these tasks are processed. We design a resource allocation and task scheduling algorithm that maximizes the rate at which tasks are processed, while considering the limits on the amount of data that can be transferred, the delay constraints and the rate at which data frames can be transmitted.}
%The resource allocation uses the computing resources available in the vehicular micro cloud and the edge servers while considering the communication constraints to maximize the rate at which computing tasks are processed. 

%The performance of our computational offloading scheme is measured at different levels of V2V penetration. 
%The results obtained in the simulated environment show that with hybrid offloading, the perception frames are processed at a significantly faster rate compared to the case with no offloading or the case with only vertical offloading.  

%The rest of this paper is organized as follows. In Section \ref{sec:sys_model}, we describe the model of the system that we assume in the development of our algorithm. The Section \ref{sec:framework} outlines our design goals and the details of the proposed algorithm. In the Section \ref{sec:results} we report the main results from the testing of our approach via simulation. The Section \ref{sec:conclusions} provides the main conclusions of this work.    

\section{System model}
\label{sec:sys_model}

\begin{figure}[t]
	\centering
	\includegraphics[width=0.35\textwidth]{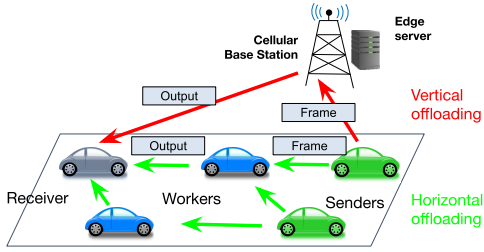}
	\caption{Overview of the system model.}
	\label{fig:sys}
	\vspace{-15pt}
\end{figure}

In this section, we describe the model of vehicular micro cloud and edge computing resources and the model of computing tasks for cooperative perception used in the remainder of this paper. 

\subsection{Vehicular and edge computing resources}
We consider a segment of a road or an intersection occupied by a set of cars $\mathcal{C}$, as illustrated in Fig. \ref{fig:sys}. A subset of cars $\mathcal{S}\in\mathcal{C}$ act as senders sharing their sensor data with other vehicles on the road. Each sender has a receiver set of cars $\mathcal{R}_s \in \mathcal{C}$, $\forall s\in \mathcal{S}$, interested in the data from the senders. We assume that all senders and receivers are V2V capable. Furthermore, the senders and receivers, together with other V2V capable cars form a stationary vehicular micro cloud $\mathcal{C}'$. A stationary micro cloud is formed by cars that are present in particular fixed geographical area \cite{hagenauer2017vehicular}. {The cars can enter and exit the area at any time and therefore the set $\mathcal{C}'$ changes over time.} On the other hand, the set of edge servers is constant over time. 

Senders and receivers can connect to a set of edge servers $\mathcal{B}$ via LTE connection through the nearby base stations (BSs). 
While we assume that all cars have cellular communications capability, only a fraction of the vehicles is capable of V2V communication, based on the current trends in car industry.

We define a set of worker nodes $\mathcal{W} \in \mathcal{B} \cup \mathcal{C}'$ that can be utilized for offloading of the processing of perception frames from the senders. The cars that take on the role of workers have computing resources available that can be used for processing of cooperative perception frames. We assume that there exist some form of incentive mechanism or agreement between car manufacturers that would lead to sharing of computing resources and sensor data between cars. The sets $\mathcal{S}$, $\mathcal{W}$ and $\mathcal{R}_1,...,\mathcal{R}_{|\mathcal{S}|}$ can be overlapping, i.e., the cars may have multiple roles from amongst the sender, receiver and worker role at the same time.

We denote the computational power, measured in Hz, of a node $i$ (either a car or an edge server) as $F_i$. Furthermore, we denote the communication rate between two nodes, $i$ and $j$, as $R_{i,j}$. We use an indicator variable $D_{i,j}^{\text{LTE}} \in \{0,1\}$, that is equal to 1 if communication between nodes $i$ and $j$ is cellular, i.e. one node of the nodes is a car and other is an edge server. Another indicator variable, $D_{i,j}^{\text{V2V}} \in \{0,1\}$, is used to capture whether the communication between the nodes $i$ and $j$ is V2V, i.e., both of the nodes are vehicles. Only $D_{i,j}^{\text{V2V}}$ or only $D_{i,j}^{\text{LTE}}$ can be equal to 1 for any given pair of nodes. 

\subsection{Computing tasks for cooperative perception}
A cooperative perception system relies on a variety of computing tasks that need to be performed before the sensor data from the senders can become useful to the receivers. Depending on the type of the sensor data, there are different types of computing tasks that might need to be performed. For example, on 3D data, generated by radar or lidar, feature extraction and localization needs to be performed using simultaneous localization and mapping (SLAM) algorithms at either the sender (the source of sensor data) or receiver cars (cars interested in the information provided by the sender). 
%Other computing tasks for cooperative perception related car pose estimation and merging of 3D frames. 
For image data, the computing tasks include feature detection and perspective transformation. We describe a computing task $l$ by a tuple of parameters $S_l = \{d_l, c_l, s_l, \mathcal{R}_l, t_l, \tau_l\}$, where:
\begin{itemize}
    \item $d_l$ is the data size of the sensor frame to be processed
    \item $c_l$ is the computational load measured in CPU cycles per second
    \item $s_l$ is the source sender of the task 
    \item $\mathcal{R}_l$ is the set of receiver nodes that are interested in receiving the output of the task
    \item $t_l$ is the time instance when the sensor frame is captured
    \item $\tau_l$ is the maximum delay. The delay is measured from $t_l$ until the task is processed and its output delivered to the receiver
\end{itemize}
We assume that every task $l$ belongs to a particular task type (TT) $k\in\{1,...,K\}$ that has a unique set of parameters $A_k = \{d_k, c_k, s_k, \mathcal{R}_k, \tau_k \}$. Therefore, for every task $l$, $ \{d_l, c_l, s_l, \mathcal{R}_l, \tau_l\} \in \left\{ A_k:k=1,...,K \right\}$. For example, a TT can be feature extraction on a 3D point cloud coming from a particular sender $s_k$, with a delay tolerance $\tau_k$, input frame size {$d_l$} and a receiver set $\mathcal{R}_k$. {Computationally demanding tasks can create a bottleneck in how many frames per second can be processed and shared between the sender and receiver cars.} {Assuming that computation is the bottleneck, the senders will generate sensor frames at the rate at which they can be processed. The sensor generation rate determines a TT arrival rate, and therefore task arrival time.}

\section{Proposed framework}
\label{sec:framework}
In this section, we propose an optimization framework for offloading of vehicular computing tasks to the edge and the vehicular micro cloud. 
%The objective is to maximize the rate at which computing tasks are performed on sensor frames, which determines the rate at which the cooperative perception systems can operate. 

{At the beginning of each period, there is a demand for completion of certain set of TTs, $\left\{ A_k:k=1,...,K \right\}$.}
{We maximize the number of tasks/frames of all TTs that are processed over the next period $T$.} Our algorithm is a two step algorithm. In the first step, we assign the computation resources $\{F_w:w \in \mathcal{W}\}$ and communication resources $\{R_{i,j}:i,j \in \mathcal{C} \cup \mathcal{B},~i \neq j \}$ to particular TTs. The resource assignment determines how many tasks of each TT will be processed at each individual worker. {Assuming that computation is the bottleneck, the senders will generate data frames to meet the achieved processing rate. In the second stage, we schedule data frame/task generation and assign where each task is processed since several workers may be processing the tasks of the same type.}  

{The optimization is done in a centralized manner. We assume that one of the cars in the vehicular micro cloud, potentially the micro cloud leader, or one of the edge servers performs the resource assignment and task scheduling based on the knowledge of communication rates $R_{i,j}$ and computing powers $F_i$ of cars and edge servers.
Since the state of a traffic scenario changes over time, the offloading decisions need to be updated periodically. Our algorithm is applied periodically at interval $T$ to obtain the resource assignment and scheduling, which is then shared with all nodes.}

\subsection{Resource assignment}
We pose the resource assignment problem as a non-linear optimization problem and then approximate it as a mixed integer linear optimization problem.

In the resource assignment stage, we solve for three optimization variables:
\begin{itemize}
    \item $X_{k,w}\in [0,1]~\forall k,w$, that determines the share of computing resources allocated by each worker $w$ for completion of tasks of type $k$
    \item $Y^\text{LTE}_{s,k,w}\in [0,1]~\forall s,k,w$, that determines the share of time resources the sender $s$ will spend on the transmission of data frames of TT $k$ to worker $w$ via cellular network
    \item $Y^\text{V2V}_{s,k,w}\in [0,1]$, that, similarly to $Y^\text{LTE}_{s,k,w}$, determines the allocation of V2V communication resources.
\end{itemize}

The delay of processing a task of a certain type $k$ needs to be less than $\tau_k$. We assume that only two types of delay are significant: the delay of transmitting the data frame to the worker if a task computation is offloaded and the delay of task processing. The data size of the task output is assumed to be negligible. {The maximum rate of transmission between a sender of TT $k$, $s_k$, and a worker $w_k$ is $R_{s_k,w}$. However, given that the sender only spends a fraction of time, $Y_{s_k,k,w}$, transmitting that particular TT to worker $w$, the effective rate of transmission is $R_{s_k,w}Y_{s_k,k,w}$. The delay of transmitting a data frame of a task of type $k$ to worker $w$ is $\frac{d_k}{R_{s_k,w}Y_{s_k,k,w}}$. The delay of processing a task at worker $w$ is $\frac{c_k}{F_wX_{k,w}}$, where $F_wX_{k,w}$ is the effective computing rate of a TT $k$ at $w$.} We define the resource assignment problem as a non-linear program: 

\begin{equation}
    \max_{{X_{k,w}, Y^\text{LTE}_{s,k,w}, Y^\text{V2V}_{s,k,w}}} {\sum_{k}\sum_{w}\left\lfloor \frac{TF_{w}X_{k,w}}{c_{k}}\right\rfloor}
    \label{eq:obj1}
    \tag{P1}
\end{equation} 
$$
\text{s.t.}
$$
\vspace{-25pt}
\begin{multline}
d_{k}\left\lfloor \frac{TF_{w}X_{k,w}}{c_{k}}\right\rfloor \leq TR_{s_{k},w}\big(D_{s_{k},w}^{LTE}Y_{s_{k},k,w}^{LTE}+\\D_{s_{k},w}^{V2V}Y_{s_{k},k,w}^{V2V}\big)
\label{eq:con1}
\tag{C1}
\end{multline} 

\begin{equation}
\sum_{k}\sum_{w}D_{s_{k},w}^{LTE}d_{k}\left\lfloor \frac{TF_{w}X_{k,w}}{c_{k}}\right\rfloor \leq U^{\text{LTE}}T
\label{eq:con2}
\tag{C2}
\end{equation} 

\begin{equation}
\sum_{k}\sum_{w}D_{s_{k},w}^{V2V}d_{k}\left\lfloor \frac{F_{w}X_{k,w}}{c_{k}}\right\rfloor \leq U^{\text{V2V}}T
\label{eq:con3}
\tag{C3}
\end{equation} 

\begin{equation}
\frac{d_{k}}{R_{s_{k},w}Y_{s_{k},k,w}^{V2V}}+\frac{c_{k}}{F_{w}X_{k,w}}\leq\tau_{k}\text{ if }X_{k,w}\neq0\text{ and }D_{s_{k},w}^{V2V}=1
\label{eq:con4}
\tag{C4}
\end{equation} 

\begin{equation}
\frac{d_{k}}{R_{s_{k},w}Y_{s_{k},k,w}^{LTE}}+\frac{c_{k}}{F_{w}X_{k,w}}\leq\tau_{k}\text{ if }X_{k,w}\neq0\text{ and }D_{s_{k},w}^{LTE}=1
\label{eq:con5}
\tag{C5}
\end{equation} 

\begin{equation}
\sum_{k}X_{k,w}\leq1,~\sum_{w}\sum_{k}Y_{s_{k},k,w}^{LTE}\leq1,~\sum_{w}\sum_{k}Y_{s_{k},k,w}^{V2V}\leq1
\label{eq:con6}
\tag{C6, C7, C8}
\end{equation}

The objective function counts the total number of tasks or data frames that will be processed over the current period $T$. The floor function $\lfloor \cdot \rfloor$ rounds its argument down to the nearest integer, since only the whole number of tasks that are processed is important. 

The Constraint (\ref{eq:con1}) is applied per each TT $k$ and it limits the rate at which data frames can be transmitted by the effective communication rate $R_{s_{k},w}Y_{s_{k},k,w}^{LTE}$ if the communication is cellular or the rate $R_{s_{k},w}Y_{s_{k},k,w}^{V2V}$ if the communication is V2V. The Constraint (\ref{eq:con2}) limits the amount of data transmitted through the cellular network by an upper bound $U^{\text{LTE}}T$ and the Constraint (\ref{eq:con3}) limits the amount of data transmitted through V2V by an uppper bound $U^{\text{V2V}}T$. 
The parameters $U^{\text{LTE}}$ and $U^{\text{V2V}}$ are defined by the {operator of the system} to precisely limit the amount of LTE and V2V traffic. %$U^{\text{LTE}}$ limits the amount of LTE traffic due to financial costs of LTE communication. Limiting the V2V communication using the bound $U^{\text{V2V}}$ is important in the case of DSRC V2V communication where the channel is shared and can become congested.

The constraints (\ref{eq:con4}) and (\ref{eq:con5}), applied per each TT $k$, are the delay constraints and they only need to be satisfied for workers $w$ that process tasks of type $k$, i.e., $X_{k,w} \neq 0$. The delay constraint (\ref{eq:con4}) applies to data frames that are transmitted via V2V communication and the delay constraint (\ref{eq:con5}) applies to transmissions that occur over cellular. 
The Constraint (C6) is applied per each worker and ensures that shares of all computing resources per worker add up to one. {Similarly, the constraints (C7) and (C8) are applied per each unique sender $s$ and they ensure that the shares of transmission time per sender add up to 1.} 

To the best of authors' knowledge, the optimization problem we arrive at cannot be readily solved by any standard optimization techniques, therefore we linearize it by making the necessary approximations. Once the optimization problem is linearized, it can be readily solved by using any mixed integer linear programming solver. 

\subsubsection*{Linearization of the optimization problem} 
A non-linear expression that appears several times in our problem formulation is $\left\lfloor \frac{TF_{w}X_{k,w}}{c_{k}}\right\rfloor $. To approximate this expression by a linear function, we introduce an integer variable $V_{k,w} \in \mathbb{Z}^+$. We add two additional constraint sets:
\begin{equation*}
    V_{k,w} \leq \frac{TF_{w}X_{k,w}}{c_{k}} \\
\end{equation*}
\begin{equation*}
    \frac{TF_{w}X_{k,w}}{c_{k}} - 0.999 \leq V_{k,w}  \\
\end{equation*}
We then replace the expression $\left\lfloor \frac{TF_{w}X_{k,w}}{c_{k}}\right\rfloor $ by the variable $V_{k,w}$ everywhere in problem formulation. 

Next, we linearize the delay constraints (\ref{eq:con4}) and (\ref{eq:con5}). We introduce a set of helper constants $\alpha^{(n)}=n/N$, for $n=0,...,N$, where $N$ is a positive integer hyperparameter. We also introduce a set of helper binary variables $U^{(n)}_{k,w} \in \mathbb{Z}_2$, $\forall n,k,w$. To simplify the exposition, we only show how one of the delay constraints can be linearized. To accomplish our goal, we introduce the following constraints that will replace the delay constraint
\begin{equation*}
    \left (1-U_{k,w}^{(n)}\right )-\alpha^{(n)}\tau_{k}\frac{F_{w}}{c_{k}}X_{k,w}\leq0 ~\forall n
\end{equation*}
\begin{equation*}
    \left(1-U_{k,w}^{(n)}\right)-\left(1-\alpha^{(n)}\right)\tau_{k}\frac{R_{s_{k},w}}{d_{k}}Y^{\text{LTE}}_{s_{k},k,w}\leq0 ~\forall n
\end{equation*}
\begin{equation*}
U_{k,w}^{(1)}+U_{k,w}^{(2)}+\dots+U_{k,w}^{(N)}\leq N-I(X_{k,w}>0)
\end{equation*}
where $I(X_{k,w}>0)$ is the indicator function equal to 1 if $X_{k,w}>0$ and 0 otherwise. This indicator function can easily be linearized using the same approach as with the expression $\left\lfloor \frac{TF_{w}X_{k,w}}{c_{k}}\right\rfloor $. The greater the value of hyperparameter $N$ selected, the more accurate our approximation becomes. 

\subsection{Task scheduling}
The resource assignment determines how many tasks will be processed by each worker over the current period $T$. Given $X_{k,w}$, the number of tasks of type $k$ processed by worker $w$ is $M_{k,w} = \left\lfloor \frac{TF_{w}X_{k,w}}{c_{k}}\right\rfloor $ and the total number of tasks of type $k$ that will be processed is $ L_k = \sum_w M_{k,w}$. {Given that workers can process $L_k$ tasks of a particular TT $k$ and assuming that that the computation is the bottleneck,
the arrival rate of sensor frames and, hence, the tasks will meet the processing rate.} Expressed mathematically, the arrival times of tasks $l=1,...,L_k$ are $t_{l}=\frac{\left(t_{l}-1\right)T}{L_k}$. The assignment of tasks to workers is performed using a heuristic policy. We use a round-robin assignment algorithm that we empirically established to minimize the queuing delays in transmission queue at the sender and the processing queue at the worker. Let $Z_{l,w}\in \mathbb{Z}_2$ be an assignment variable equal to 1 if the task $l$ is assigned to worker $w$ and 0 otherwise. The heuristic round robin assignment policy is described in the pseudo-code below:
\begin{algorithm}[]
\DontPrintSemicolon
\SetAlgoLined
\KwResult{$Z_{l,w}$}
$Z_{l,w} \leftarrow 0~\forall w \in \mathcal{W}, \forall l=1,...,L_k$\;
$M_{k,w} \leftarrow \left\lfloor \frac{TF_{w}X_{k,w}}{c_{k}}\right\rfloor ~\forall w \in \mathcal{W}$\;
$ L_k \leftarrow \sum_w M_{k,w}$\;
$l \leftarrow 0$\;
\While{$l<L_k$}{
    \For{$w \in \mathcal{W}$}{
        \If{$M_{k,w} \neq 0$}{
            $Z_{l,w} \leftarrow 1$\;
            $M_{k,w} \leftarrow M_{k,w} - 1$\;
            $l \leftarrow l + 1$
        }
    }
}

\caption{Round-robin task assignment for a TT $k$}
\end{algorithm}
The policy is applied per each TT separately. The algorithm continuously loops over each of the workers that are sorted randomly and assigns each of the $L_k$ tasks to one worker in each loop iteration unless that worker has already been assigned $M_{k,w}$ tasks. The looping over workers ends once all $L_k$ tasks have been assigned.   
\section{Simulation results}
\label{sec:results}

% \begin{figure}[t]
% 	\centering
% 	\includegraphics[width=0.15\textwidth]{figures/intersection.png}
% 	\caption{The intersection simulated in the LUST scenario. The micro cloud is formed by the cars that enter the circle of radius 150 m shown in the figure.}
% 	\label{fig:lust_scenario}
% \end{figure}

In this section, we analyze the performance of the proposed approach for horizontal and vertical computing offloading for cooperative perception in a simulated traffic environment. We simulate an intersection in the Luxembourg SUMO traffic (LuST) scenario \cite{codeca2015luxembourg}. The simulated intersection is located at 49°36'34.7"N, 6°07'09.7"E and the micro cloud area radius is 150 m. We assume that all V2V capable cars that enter the circular area become members of the vehicular micro cloud. The traffic is simulated for one hour between 8 AM and 9 AM. 

A share $\eta_S$ of cars in the micro cloud are randomly selected to be senders and a share $\eta_R$ are randomly selected to be receivers. All the cars in the micro cloud, including the senders, together with one edge server form the set of workers.

{The cars communicate to each other via DSRC V2V communication while the communication to the edge server is done via LTE. In our simulations, we model the LTE and V2V communication links as pipes with some rate $R_{i,j}$. In the LUST scenario, using Veins \cite{sommer2010bidirectionally}, we simulate the cars broadcasting beaconing signals via DSRC three times per second, which we use to estimate the signal to noise plus interference ratio (SINR) between cars in the area. The SINR is mapped to DSRC communication rates based on the mappings reported in \cite{jiang2008optimal}. The LTE communication rate to the edge server is modeled as a Gaussian random variable $\mathcal{N}\left(\mu_R, \sigma_R \right)$. We assume that the knowledge of the communication rates $R_{i,j}$ is available to the node that calculates the optimal resource assignment. In practice, these values would need to be obtained via some distributed or centralized rate prediction algorithm} 

The computing power $F_i$ of sender cars is $\mu_C^{(1)}$. From the remaining cars in the micro cloud, 70\% also have the computing power $\mu_C^{(1)}$ while the remainder of the cars have the computing power $\mu_C^{(2)}$ {, where $\mu_C^{(2)} > \mu_C^{(1)}$. The computation power $\mu_C^{(2)}$ corresponds to high-end vehicles that have abundant computation power compared to regular vehicles.} Finally, the computing power of the edge server is $\mu_C^{(3)}$. {We assume that all nodes report their computing power to the node in charge of performing resource assignment.} 

We {assume} that there is one TT per sender per each optimization period and that all of the TTs have the parameters $d_k$, $c_k$ and $\tau_k$ in common. We obtain a reference for the values of these parameters from some reported traces of computer vision computing tasks \cite{kattepur2017priori}. The simulations are performed for two generic types of computing tasks, one representing image processing tasks such as edge detection, and one representing {radar}  point cloud processing tasks such as a SLAM operation. The point cloud type has a higher computational load $c_k$ than the image type but a lower data size $d_k$. {The maximum latency $\tau_k$ for sensor frame processing can not be too high as the positions of the vehicles on the road change rapidly}. The values of parameters used in simulations are given in Tables I and II.

{As a benchmark for our resource assignment algorithm, we use random resource assignment. With random resource assignment, all senders are assigned to process their own TTs and each of the remaining workers and the edge servers are randomly assigned to a TT at maximum capacity. The transmission resources $Y_{s_k,k,w}$ at each sender are equally split across the workers that process its TTs. Naturally, with this random assignment approach we cannot guarantee that the constraints (C1-C5) will be satisfied.}

\begin{table}[t]
\vspace{5pt}
\renewcommand{\arraystretch}{1.3}
\caption{Computing and communications resources and parameters used in simulations}
\label{table:params}
\centering

\begin{tabular}{c||c||c||c}
\hline 
Parameter & Value & Parameter & Value\tabularnewline
\hline 
\hline 
$T$ & 1 s & $\eta_S$ & 0.2\tabularnewline
\hline 
$[\mu_C^{(1)}, \mu_C^{(2)}, \mu_C^{(3)}]$ & [1 5 10] GHz & $\eta_R$ & 0.2\tabularnewline
\hline 
$\mu_R$ & 50 Mb/s & $U^{\text{V2V}}$ & $\infty$\tabularnewline
\hline 
$\sigma_R$ & 5 Mb/s & $U^{\text{LTE}}$ & 24 Mb/s\tabularnewline

\hline
\end{tabular}

\end{table}

\begin{table}[t]
\renewcommand{\arraystretch}{1.3}
\caption{Parameter of image and point cloud processing tasks used in simulations}
\label{table:params}
\centering

\begin{tabular}{c|c|c|c}
\hline 
Parameter & $c_k$ & $d_k$ & $\tau_k$\tabularnewline
\hline 
Image & 1E9 & 20 KB & 0.6 s\tabularnewline
\hline 
Point cloud & 2E8 & 400 KB & 0.6 s\tabularnewline
\hline 

\end{tabular}
\vspace{-10pt}
\end{table}

\subsection{Image processing}

\begin{figure}[t]
\centering
\begin{subfigure}{.24\textwidth}
  \centering
  \includegraphics[width=0.9\linewidth]{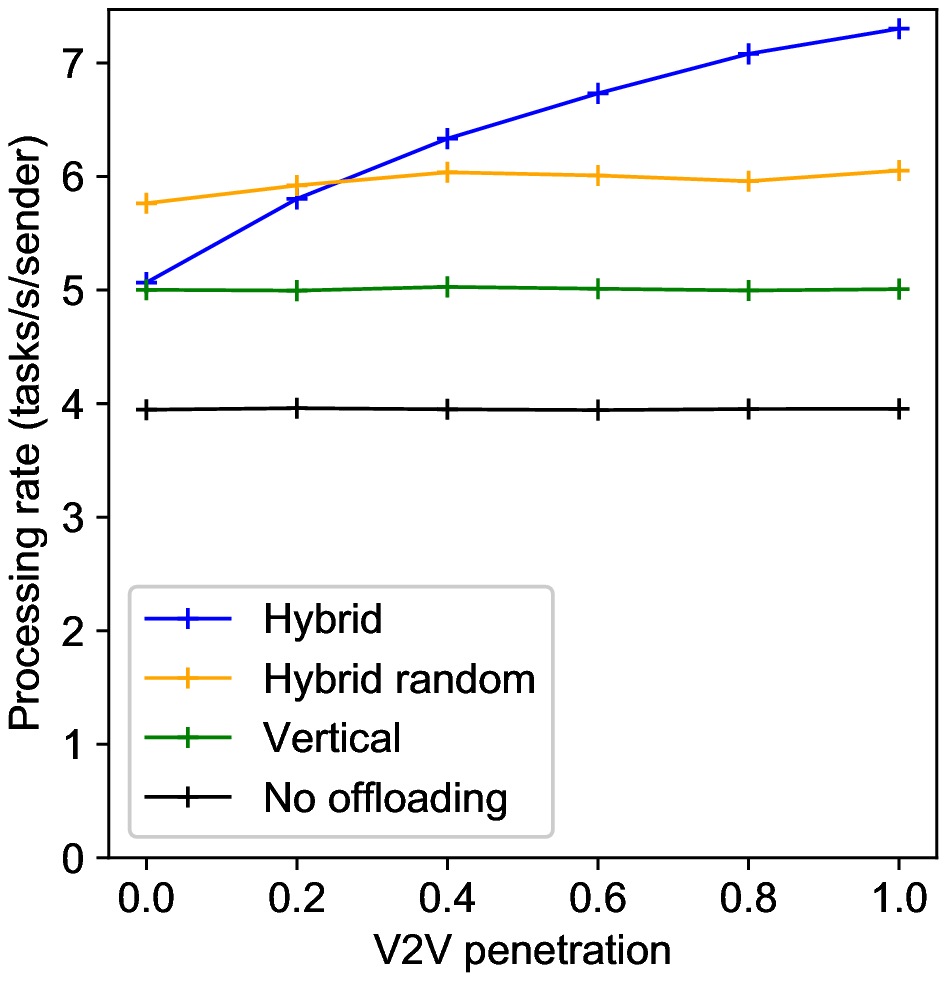}
  \caption{With LTE cap}
  \label{fig:rate_image_no_cap}
\end{subfigure}%
\begin{subfigure}{.24\textwidth}
  \centering
  \includegraphics[width=0.9\linewidth]{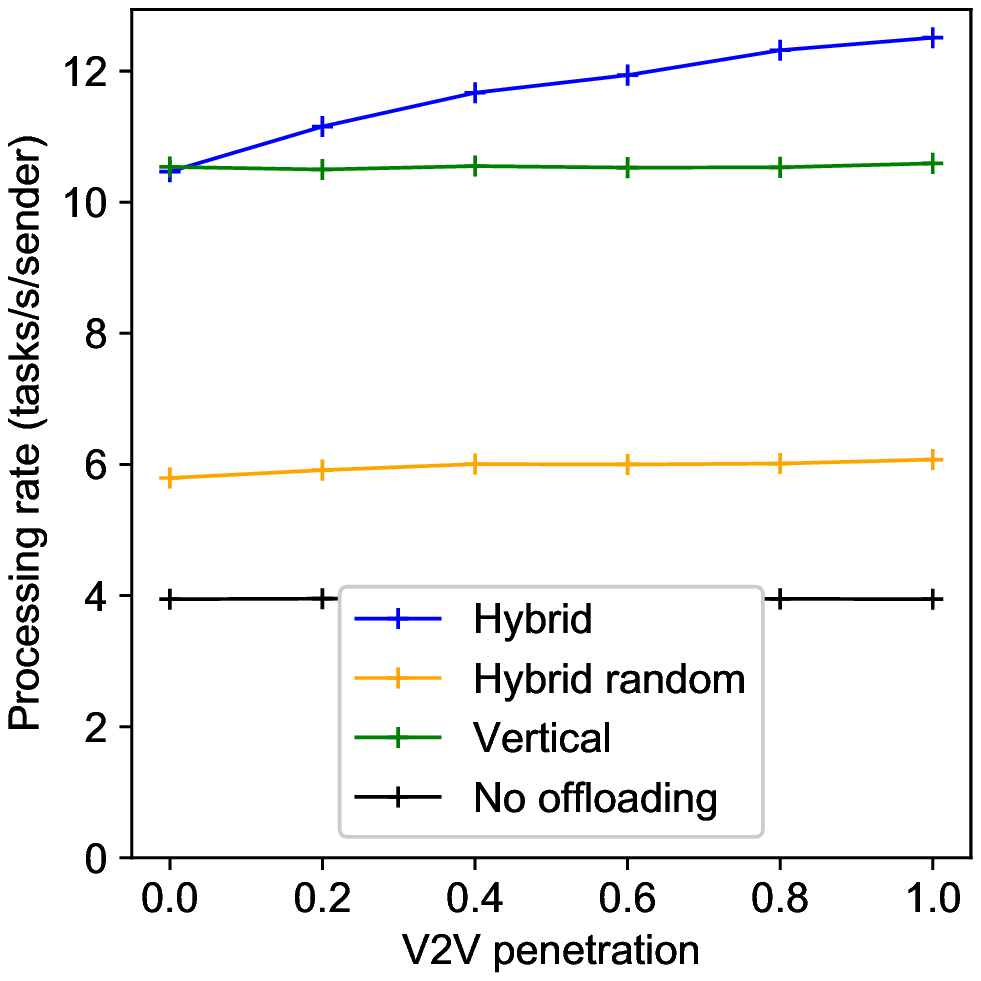}
  \caption{Without LTE cap}
  \label{fig:rate_image_with_cap}
\end{subfigure}
\caption{The processing rate of image tasks.}
\label{fig:rate_image}
\vspace{-5pt}
\end{figure}

We first analyze and compare the offloading approaches for image processing tasks. The performance metric is the number of tasks per second per sender that are processed within the allowed latency $\tau_k$. The results are shown in Fig. \ref{fig:rate_image}. with a limit $U^{\text{LTE}}  = 24$ MB/s on the amount of LTE transmissions and without any limit on LTE traffic. 

{The LTE uplink limit can restrict the number of tasks that can be offloaded to the edge over a period $T$. The results in Fig. \ref{fig:rate_image_no_cap} demonstrate what the performance may look like if the uplink limit becomes the bottleneck. Vertical offloading does not provide a significant benefit in terms of the processing rate. However, horizontal offloading with the help of vertical offloading (\emph{hybrid offloading}) can still double the processing rate compared to the case of no offloading.} The number of tasks processed by horizontal offloading is unaffected by the LTE cap but it does depend on the V2V penetration defined as the share of cars other than senders and receivers that are equipped for V2V communication and hence able to serve as workers. {Without an LTE data transmission limit, vertical offloading doubles the processing rate on its own. However, this scenario is unrealistic as it does not recognize the financial cost that can be incurred due to LTE upload.}

{We observe that with hybrid offloading and random assignment the system resources are underutilized, and the performance only slightly increases with higher penetration. When tasks are offloaded to randomly selected vehicles, the communication rate to the selected vehicles may not always be sufficient to deliver the frames fast enough and so less frames are successfully processed. However, we should note that this is not a one-to-one comparison to our algorithm since with random assignment the uplink limit constraint is not always satisfied, which is why with no V2V penetration the random assignment seemingly performs better in Fig. 2a.} 

\begin{figure}[t]
	\centering
	\includegraphics[width=0.35\textwidth]{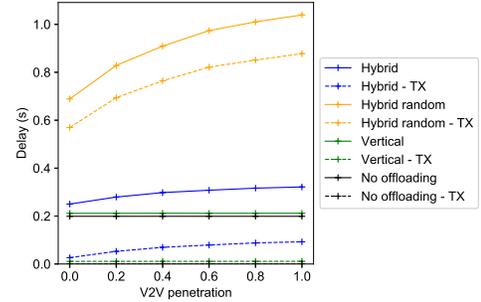}
	\caption{The average processing delay, which consists of transmit and computing delay, for image tasks. The transmit delay is represented by the dotted line.}
	\label{fig:image_delay}
	\vspace{-15pt}
\end{figure}

We also analyze how offloading affects the average delay of the processing of the frames. The results are shown in Fig. \ref{fig:image_delay} for the case with an LTE traffic limit. The delay consists of transmission delay to the processor, if a task is offloaded, and the compute delay at the processor. We assume that the transmission delay of the output results is negligible. The average transmission delay of hybrid offloading is higher than that of vertical offloading because a DSRC connection, which is utilized for horizontal offloading, has a lower rate than an LTE connection. The transmission delay increases with V2V penetration since larger share of tasks is offloaded horizontally therefore the average transmission delay is larger. The average compute delay (the transmission delay subtracted from the total delay) moderately decreases with V2V penetration since some tasks are being offloaded to the high-end vehicles. Overall, offloading image tasks is only suitable in the scenarios where it is acceptable to incur an additional processing delay in order to achieve a higher rate. {With random assignment, the transmission delay is very significant because data frames spend extensive amount of time in the transmission queue since the transmission rate cannot satisfy the scheduled rate.}

\subsection{Point cloud processing}

\begin{figure}[t]
	\centering
	\includegraphics[width=0.25\textwidth]{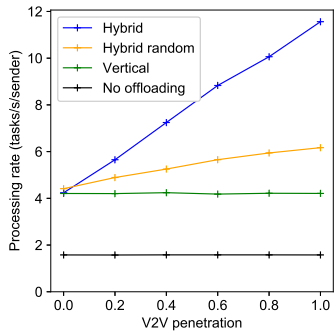}
	\caption{The processing rate of point cloud tasks. }
	\label{fig:rate_point_cloud}
	\vspace{-15pt}
\end{figure}

\begin{figure}[t]
	\centering
	\includegraphics[width=0.35\textwidth]{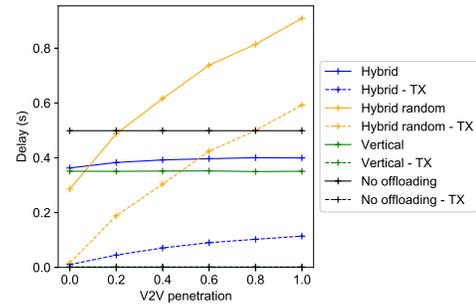}
	\caption{The average processing delay, which consists of transmit and computing delay, for point cloud tasks.}
	\label{fig:delay_point_cloud}
	\vspace{-15pt}
\end{figure}

The processing rate for point cloud tasks is shown in Fig. \ref{fig:rate_point_cloud}. Since the data size of point cloud frames is small, the LTE traffic is already below the cap that we set, and it does not have an impact on the processing rate. Vertical offloading increases the processing rate by around 100\% in our scenario. Since the transmission overhead is smaller than that of image tasks, large gains are possible thanks to horizontal offloading and, overall, multiple-fold gain is obtained with hybrid offloading. {Point cloud tasks lend themselves much better to offloading compared to image tasks. This is further supported by the delay profiles shown in Fig. \ref{fig:delay_point_cloud}. Since the compute load of point cloud tasks is significantly larger than that of the image tasks, their processing delay can be significantly reduced by offloading them to more powerful processors.} Indeed, we observe a significant decrease in the processing delay with both vertical and hybrid offloading in Fig. \ref{fig:delay_point_cloud}. due to offloading to the edge server and to the powerful high-end cars. Overall, offloading of point cloud tasks increases the processing rate while decreasing the processing delay.

% \begin{figure}[t]
% \centering
% \begin{subfigure}{.24\textwidth}
%   \centering
%   \includegraphics[width=1\linewidth]{figures/process_rate_vs_v2v_penetration_image2.png}
%   \caption{Point cloud processing}
%   \label{fig:sub1}
% \end{subfigure}%
% \begin{subfigure}{.24\textwidth}
%   \centering
%   \includegraphics[width=1\linewidth]{figures/process_rate_road.pdf}
%   \caption{Image processing}
%   \label{fig:sub2}
% \end{subfigure}
% \caption{The achieved processing rate in two different scenarios.}
% \label{fig:rate}
% \end{figure}

% \begin{figure}[t]
% \centering
% \begin{subfigure}{.24\textwidth}
%   \centering
%   \includegraphics[width=1\linewidth]{figures/delay_vs_v2v_penetration_intersection.pdf}
%   \caption{Intersection}
%   \label{fig:sub1}
% \end{subfigure}%
% \begin{subfigure}{.24\textwidth}
%   \centering
%   \includegraphics[width=1\linewidth]{figures/delay_vs_v2v_penetration_road.pdf}
%   \caption{Road}
%   \label{fig:sub2}
% \end{subfigure}
% \caption{The delay of task processing in two different scenarios.}
% \label{fig:delay}
% \end{figure}

% \begin{figure}[t]
% \centering
% \begin{subfigure}{.24\textwidth}
%   \centering
%   \includegraphics[width=1\linewidth]{figures/loss_vs_v2v_penetration_intersection.pdf}
%   \caption{Intersection}
%   \label{fig:sub1}
% \end{subfigure}%
% \begin{subfigure}{.24\textwidth}
%   \centering
%   \includegraphics[width=1\linewidth]{figures/loss_vs_v2v_penetration_road.pdf}
%   \caption{Road}
%   \label{fig:sub2}
% \end{subfigure}
% \caption{The task loss in two different scenarios.}
% \label{fig:loss}
% \end{figure}

\section{Conclusions}
\label{sec:conclusions}
%Vehicular cooperative perception is a promising paradigm in which cars share their perception sensor information with their neighbors helping them to expand their observed area and gain a more complete picture of the traffic. 
In vehicular computing-intensive applications, the task processing rate can be increased by offloading the computing to the local edge servers (vertical offloading) and to the nearby cars (horizontal offloading). We develop an optimized resource assignment and scheduling algorithm for hybrid offloading of computing tasks for cooperative perception that maximizes the rate at which frames are processed, while ensuring that results are delivered within a deadline and also constraining the cellular and V2V communication overhead. 
The algorithm is tested in a simulated environment {based on the LUST traffic scenario and Veins vehicular network simulator.} We observe a significant increase in the processing rate of sensor frames when using hybrid offloading compared to the no offloading case or the case with only vertical offloading.

\bibliography{references}
\bibliographystyle{ieeetr}

\end{document}